\documentclass[runningheads]{llncs}
\usepackage[T1]{fontenc}
\usepackage{graphicx}
\usepackage{amsmath}
\usepackage[colorlinks=true, linkcolor=blue, citecolor=blue, urlcolor=blue]{hyperref}
\usepackage{multicol, multirow}
\usepackage{booktabs} 
\usepackage{amsmath,amssymb,xcolor,fontawesome5} 

\usepackage{footmisc}

\begin{document}
\title{Frequency-Aware Ensemble Learning for BraTS 2025 Pediatric Brain Tumor Segmentation}
%
%

\author{Yuxiao Yi\inst{1}\orcidID{0009-0002-6409-2760} \and
        Qingyao Zhuang\inst{1}\orcidID{0009-0004-5754-4727} \and 
        Zhi-Qin John Xu\inst{1,*}\orcidID{0000-0003-0627-3520}\and
        Xiaowen Wang\inst{3,*} \orcidID{0009-0004-5559-0981}\and
        Yan Ren\inst{2,*}\orcidID{0000-0001-5993-9248}\and
        Tianming Qiu\inst{3,*}\orcidID{0000-0003-1089-4717}}

\authorrunning{Y. Yi et al.} 
\institute{
 \textsuperscript{1} Institute of Natural Sciences, School of Mathematical Sciences, MOE-LSC, Shanghai Jiao Tong University, Shanghai, China \\
 \email{\{yiyuxiao,alan\_zqy,xuzhiqin\}@sjtu.edu.cn} \\
 \textsuperscript{2} Department of Radiology, Huashan Hospital, Fudan University, Shanghai, China \\
\email{renyan\_richard@aliyun.com} \\
 \textsuperscript{3} Department of Neurosurgery, Huashan Hospital, Fudan University, Shanghai, China \\
\email{ \{apolloslisy,tianming2100\}@126.com} \\
} 
\maketitle              
%

\let\origfootnote\thefootnote
\renewcommand{\thefootnote}{}
\footnotetext{\textsuperscript{*} Corresponding authors.}
\let\thefootnote\origfootnote

\begin{abstract}
	Pediatric brain tumor segmentation presents unique challenges due to the rarity and heterogeneity of these malignancies, yet remains critical for clinical diagnosis and treatment planning. We propose an ensemble approach integrating nnU-Net, Swin UNETR, and HFF-Net for the BraTS-PED 2025 challenge. Our method incorporates three key extensions: adjustable initialization scales for optimal nnU-Net complexity control, transfer learning from BraTS 2021 pre-trained models to enhance Swin UNETR's generalization on pediatric dataset, and frequency domain decomposition for HFF-Net to separate low-frequency tissue contours from high-frequency texture details. 
    Our final ensemble framework combines nnU-Net ($\gamma=0.7$), fine-tuned Swin UNETR, and HFF-Net, achieving Dice scores of 62.7\% (CC), 83.2\% (ED), 72.9\% (ET), 85.7\% (NET), 91.8\% (TC), and 92.6\% (WT) on the unseen test dataset, respectively. 
    \textit{Our proposed method achieves first place (rank 1st) in the BraTS 2025 Pediatric Brain Tumor Segmentation Challenge}.

\end{abstract}
\section{Introduction}
\label{sec:intro}

Brain tumors represent one of the most severe malignancies threatening pediatric health worldwide. These tumors typically exhibit high invasiveness and poor prognosis. Multi-parametric magnetic resonance imaging (mpMRI) has become the fundamental non-invasive modality for pediatric brain tumor diagnosis, typically comprising multiple scanning sequences. However, manual segmentation relies on the expertise of clinicians or technicians, which is not only time-consuming and labor-intensive but also susceptible to inter-operator variability. Therefore, developing accurate, efficient, and robust automatic segmentation methods holds significant value for clinical diagnosis, treatment planning, and prognostic assessment.

Recent advances in deep learning have revolutionized medical image analysis, demonstrating exceptional capabilities in automating complex segmentation tasks with performance approaching or even surpassing human expert levels.
In response to these technological advances and clinical demands, the Brain Tumor Segmentation Challenge (BraTS) has expanded its scope to include pediatric cases, assembling the largest annotated pediatric brain tumor imaging dataset to date \cite{karargyris2023NMI,kazerooni2024a,ped2023,ped2024}. This initiative provides the biomedical community with invaluable resources, driving the development of automated segmentation algorithms.

Winning methods in recent years have extensively adopted the self-configuring nnU-Net \cite{isensee2021NM} as their baseline model \cite{sadique2024LNiC,jiang2024,jiang20242IIS}. Its U-shaped network architecture, based on convolutional neural networks (CNNs), comprises an encoder, decoder, and skip connections.
While CNNs possess strong feature extraction capabilities, their local receptive field characteristics limit their ability to capture long-range dependencies.
Inspired by the success of the transformer \cite{Vaswani2017} in natural language processing (NLP), researchers have integrated attention mechanisms into segmentation architectures. The
UNETR \cite{hatamizadeh20222IWC} combines Vision Transformer \cite{vit} encoders with CNN decoders, while Swin UNETR \cite{he2023LNiC,hatamizadeh2022LNiC} employs Swin Transformer \cite{liu2021swin} to achieve state-of-the-art performance. Recent innovations like SegFormer3D \cite{perera2024} use hierarchical transformers to extract multi-scale features with fewer parameters.
Meanwhile, the success of pre-trained models like SAM \cite{kirillov20232IIC} has popularized the pre-training and fine-tuning paradigm in biomedical segmentation \cite{shi2025,wang2024sammed3d,Pachitariu2025,huang2024LNiC,akbar2024LNiC,wang2025CV-E}. Models trained on large datasets are fine-tuned for specific tasks to improve performance and generalization. These technological advances provide powerful tools for automatic segmentation of pediatric brain tumors.

In this work, we propose an ensemble method combining nnU-Net, Swin UNETR, and the newly proposed HFF-Net \cite{shao2025ITMI}, trained on the BraTS-PED 2025 dataset.
Our approach includes three key extensions:
1). tunable initialization scales for nnU-Net complexity control;
2). transferring BraTS 2021 pre-trained models to BraTS-PED 2025 for improved Swin UNETR training;
3). and introducing frequency domain decomposition to separate smooth tissue contours from texture details, enhancing tumor segmentation accuracy.

\section{Method}
\label{sec:method}

In this paper, we employ nnU-Net \cite{isensee2021NM}, Swin UNETR \cite{he2023LNiC,hatamizadeh2022LNiC} and HFF-Net \cite{shao2025ITMI} as our baseline models. For nnU-Net, we modify the default initialization method and train separate models with different initialization scales. Additionally, for the transformer-based Swin UNETR model, we adopt a pre-training and fine-tuning paradigm, where a model trained on large-scale datasets serves as the backbone and is fine-tuned on downstream tasks to achieve better generalization.

\subsection{Data Description}
\label{sec:data}

The BraTS-PED Challenge aims to perform segmentation and auxiliary clinical analysis of malignant primary pediatric brain tumors. The inaugural BraTS-PED Challenge was successfully held in 2023, followed by the collection of a larger dataset in 2024 to enrich dataset diversity. In 2025, minor improvements were made, including the addition of inter-rater and intra-rater variability assessments for the test dataset annotations. This validates the consistency and reproducibility of automatic annotations, enabling the development of more robust segmentation algorithms.

Current challenge provides a dataset of 438 pediatric high-grade glioma cases, comprising 261 training cases, 91 validation cases, and 86 test cases. Participants can publicly download the training data for model development, while the test data remains private and is used exclusively for final evaluation. Each case contains four MR scanning sequences: pre-contrast native T1-weighted (T1N), contrast-enhanced T1-weighted (T1C), T2-weighted (T2W), and T2-weighted Fluid Attenuated Inversion Recovery (T2-FLAIR). To generate the corresponding segmentation masks, deep learning-based automatic segmentation models are first used to produce coarse pre-segmentation results, which are then refined by experienced annotators to obtain the final segmentation labels. The segmentation results contain five single-value labels representing the background and four tumor-related regions. Unlike previous challenges, BraTS-PED 2025 focuses on six sub-regions: enhancing tumor (ET), non-enhancing tumor (NET), cystic component (CC), peritumoral edema (ED), and two composite regions tumor core (TC) and whole tumor (WT).

\setcounter{footnote}{0}

\subsection{Data Pre-processing}
\label{sec:data_pre}

The original data has undergone only facial feature removal for privacy protection, without skull stripping. We first perform skull stripping on all training and validation data. The processed four MRI modalities are then fed into nnU-Net and Swin UNETR for training. We employ a publicly available deep learning-based pipeline \footnote{\url{https://github.com/d3b-center/peds-brain-auto-skull-strip}} to generate skull-stripping masks for pediatric patients. 
The pipeline leverages a pre-trained nnU-Net (v1) model as its backbone, takes four modalities as input, and outputs the corresponding skull mask.
During the final testing phase, each test case undergoes the same procedure to ensure consistency.
Additionally, each modality undergoes frequency domain decomposition to extract one low-frequency component and four directional high-frequency components. Low-frequency decomposition via Dual-Tree Complex Wavelet Transform (DTCWT) produces smoothed brain tissue images, while high-frequency decomposition through Non-subsampled Contourlet Transform (NSCT) captures tissue texture and directional features \footnote{\url{https://github.com/VinyehShaw/HFF}}. 
Unlike traditional wavelet transforms, the DTCWT employs a dual-tree structure with carefully timed subsampling across two independent filter banks. This design prevents the image size from halving at each level. Additionally, to ensure consistent image size, the NSCT employs a two-step decomposition process using a Non-Subsampled Pyramid (NSP) filter and a Non-Subsampled Directional Filter Bank (NSDFB) to partition a two-dimensional image. 
\begin{figure}[htbp]
	\centering
	\includegraphics[width=\linewidth]{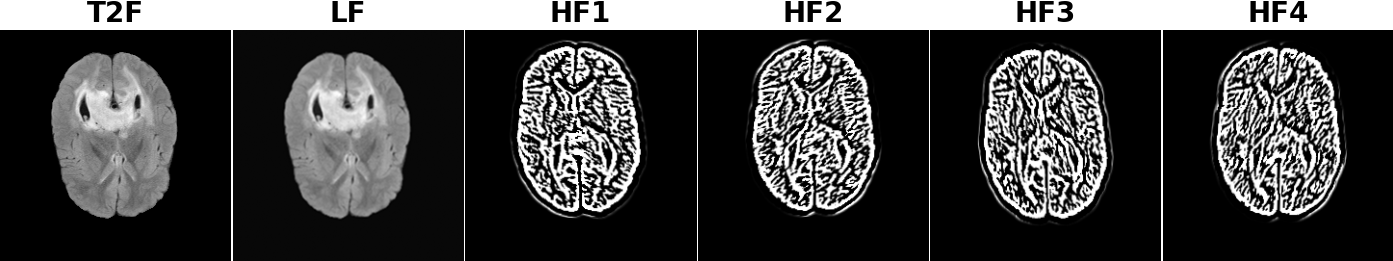}
	\caption{ \textbf{Frequency decomposition results} of T2-FLAIR on the training set. \textit{LF} denotes low frequency component and \textit{HF1}$\sim$\textit{HF4} represents four high frequency components, respectively. }
	\label{fig:HFF}
\end{figure}
\begin{figure}[htbp]
	\centering
	\includegraphics[width=\linewidth]{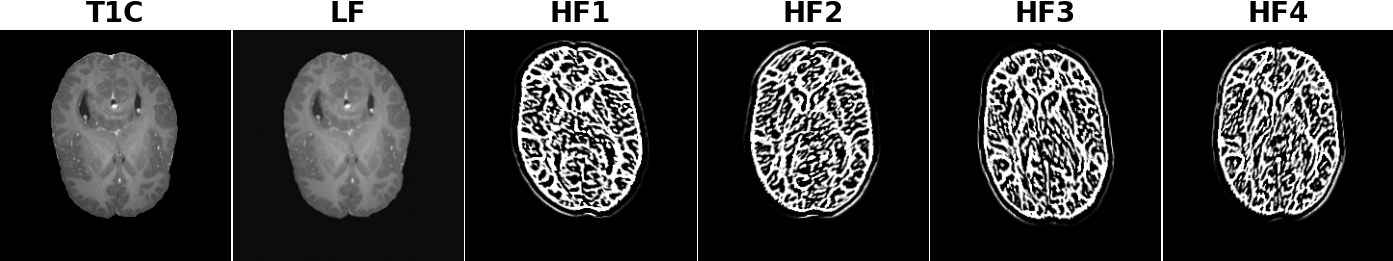}
	\caption{ \textbf{Frequency decomposition results} of T1C on the training set. \textit{LF} denotes low frequency component and \textit{HF1}$\sim$\textit{HF4} represents four high frequency components, respectively. }
	\label{fig:HFF_t1c}
\end{figure}

The nnU-Net and Swin UNETR take the 4 skull-stripped modalities as input, while HFF-Net receives the corresponding 4 low-frequency parts and 16 high-frequency modalities.


\subsection{Model Training}
\label{sec:nn}





\subsubsection{nnU-Net}
\label{sec:nnunet} is a deep learning framework based on the classical U-Net architecture, specifically designed for medical image segmentation. It automatically configures pre-processing pipelines, network architectures, training procedures, and post-processing schemes for new tasks in the biomedical domain, eliminating the need for tedious manual parameter tuning or purely empirical approaches. In biomedical segmentation tasks, nnU-Net achieves more accurate predictions and robust generalization compared to other specialized solutions, making it a commonly used baseline model in BraTS challenges.

We employ two weight initialization strategies: 1). The default nnU-Net initialization, where linear layer weights are initialized using uniform distribution and convolutional layers use Kaiming initialization; 2). All modules are initialized using normal distributions $\mathcal{N}(0, \left( \frac{1}{d^{\gamma}} \right)^2 )$, where $d$ is the number of input neurons and $\gamma$ is a tunable hyperparameter. A higher $\gamma$ value leads to smaller initialization scales \cite{zhang2025}. Based on these initialization methods, we train six 3D full resolution nnU-Net (v2) models, including one with default initialization and five with $\gamma$ values of 0.3, 0.5, 0.7, 0.9, and 1.0, respectively.

For each model, we adopt five-fold cross-validation. The original 261 cases are split into training and validation sets at an 8:2 ratio. The training batch size is 2, with an input patch cropped from 3D images of $96\times160\times160$ voxels. Models are trained for 1000 epochs, with each epoch containing 250 mini-batches. During each training step, data is randomly loaded and applied with dynamical augment strategies, including rotation, scaling, Gaussian noise, Gaussian blur, brightness adjustment, contrast adjustment, low-resolution simulation, gamma correction, and mirroring. 

The loss function is a weighted sum of Dice loss and cross-entropy loss. Stochastic gradient descent with Nesterov momentum ($\mu=0.99$) is used as the optimizer. The learning rate follows a polynomial-decay schedule, expressed as:
\begin{equation}
	lr = lr_{\text{init}} \times \left( 1- \frac{  \text{epoch} }{ \text{max\_epoch}   } \right)^{0.9} \label{eq:lr}
\end{equation}
where the initial learning rate $lr_{\text{init}} = 10^{-2}$.

\subsubsection{Swin UNETR}
\label{sec:swinunetr}

is a hierarchical Transformer-based model that replaces the encoder in the U-Net architecture with a Swin Transformer, which computes self-attention through an efficient shifted window partitioning scheme. Compared to CNNs with local receptive fields, Swin Transformer demonstrates superior capabilities in multi-scale contextual representation learning and long-range dependency modeling.

To leverage the robust data processing, training, and inference pipeline of nnU-Net, we integrate the Swin UNETR model into the nnU-Net framework, enabling out-of-the-box functionality. We first train a model from scratch on the BraTS-PED 2025 dataset as a baseline, maintaining the same training parameters as described above. Additionally, we employ a Swin UNETR model pre-trained on the BraTS 2021 segmentation challenge dataset as the foundation model and fine-tune it for our BraTS-PED segmentation task. The pre-trained model is trained for 800 epochs on 1,251 training samples with an initial learning rate of 8e-4, and the corresponding model weights are available in the GitHub repository\footnote{\url{https://github.com/Project-MONAI/research-contributions/tree/main/SwinUNETR/BRATS21}}. For the fine-tuning task, to maintain compatibility with the pre-trained model, the input patch size is set to $128\times128\times128$ voxels with deep supervision disabled. The model is trained for 1,000 epochs with a batch size of 2, following the learning rate scheduling strategy outlined in Eq.~\ref{eq:lr}, except that the initial learning rate $lr_{ \text{init}}$ is adjusted to $10^{-3}$.

\subsubsection{HFF-Net}
\label{sec:hff} introduces a novel dual-branch framework that leverages frequency-domain decomposition to enhance brain tumor segmentation, particularly for contrast-enhancing regions. The architecture initially decomposes multi-modal MRI scans into low-frequency (LF) using Dual-Tree Complex Wavelet Transform (DTCWT), and multi-directional high-frequency (HF) components with Non-subsampled Contourlet Transform (NSCT).

In this paper, the inputs of HFF-Net consist of 4 low-frequency modalities and 16 high-frequency modalities, all derived from the original 4 MRI scans via a frequency decomposition module. The model is trained for 450 epochs with a batch size of 1 and a patch size of $128\times128\times128$. We employ an SGD optimizer with a momentum of 0.9 and a weight decay of $5\times10^{-5}$. The initial learning rate is set to 0.3 and progressively decays during training. All inputs undergo Z-score normalization.

All training procedures are conducted on an NVIDIA GeForce RTX 4080 GPU with 16 GB memory. Among them, training a single fold of nnU-Net requires approximately 37 hours, while HFF-Net needs 48 hours. Swin UNETR is the most time-consuming, requiring 4 days.

\subsection{Model Ensemble}
\label{sec:ensemble}

To enhance the model's accuracy and generalization performance on unseen datasets, we employed a model ensemble strategy. We computed the weighted average of the prediction probabilities from nnU-Net, Swin UNETR, and HFF-Net, using equal weights of $1/3$ for each model. This averaged probability map was then converted back to the final segmentation mask labels. It is worth noting that we did not employ any post-processing techniques, such as feature extraction of connected components.

\subsection{Evaluation Metrics}
\label{sec:metrics}

The evaluation of the six subregions follows the BraTS Lighthouse Challenge framework using two key metrics:
\begin{itemize}
	\item Lesion-wise Dice Similarity Coefficient (DSC) measures voxel-level overlap between predicted and reference segmentations for each individual lesion, excluding true-negative voxels.

	\item Normalized Surface Distance (NSD) assesses boundary accuracy between predictions and ground truth.
\end{itemize}

\section{Results}
\label{sec:results}

\subsection{Quantitative results}
\label{sec:quantitative}

Table \ref{tab:ped-val} presents the quantitative evaluation results of our models on the BraTS-PED validation dataset. All predicted segmentation labels were obtained through five-fold cross-validation. The entire evaluation process was automatically conducted using the pipeline provided by the challenge organizers on the Synapse platform. It should be noted that participants have access only to validation images without ground truth labels, while both the test images and the corresponding labels remain confidential.

\begin{table}[h!]
    \caption{\textbf{Quantitative results} on the validation datasets of PED. Lesion-wise (LW) Dice coefficients and Normalized Surface Dice at 0.5 mm tolerance (NSD-0.5) were computed for enhancing tumor (ET), tumor core (TC), whole tumor (WT), non-enhancing tumor (NET), cystic components(CC), and edema (ED), respectively.}
    \renewcommand{\arraystretch}{1.15}  
    \centering
    \resizebox{1.0 \textwidth}{!}{%
        \begin{tabular}{@{}c l *{6}{c} c *{6}{c}@{}}
            \toprule
            \multirow{2}{*}{\textbf{Task}} & \multirow{2}{*}{~~~~\textbf{Model}} & \multicolumn{6}{c}{\textbf{Lesion-wise Dice $\uparrow$ }} & & \multicolumn{6}{c}{\textbf{Lesion-wise NSD-{0.5} $\uparrow$}} \\
            \cmidrule(lr){3-8} \cmidrule(lr){10-15}
            & & \textbf{CC} & \textbf{ED} & \textbf{ET} & \textbf{NET} & \textbf{TC} & \textbf{WT} & & \textbf{CC} & \textbf{ED} & \textbf{ET} & \textbf{NET} & \textbf{TC} & \textbf{WT} \\
            \midrule
            \multirow{10}{*}{\begin{tabular}[c]{@{}c@{}}PED\\Val.\\N=91\end{tabular}}
            & nnU-Net                    &0.657	&0.967	&0.625	&0.9	&0.931	&0.93 & & 0.648 & 0.967 & 0.549 & 0.639 & 0.661 & 0.662 \\
            & nnU-Net ($\gamma=0.3$)    &0.692	&0.945	&0.677	&0.895	&0.924	&0.925 & & 0.677 & 0.945  & 0.615  & 0.657  & 0.679 & 0.68 \\
            & nnU-Net ($\gamma=0.5$)     &0.691	&0.956	&0.671	&0.899	&0.928	&0.928 & & 0.675 & 0.956  & 0.606  & 0.663  & 0.686 & 0.686 \\
            & nnU-Net ($\gamma=0.7$)    &0.75	&0.956	&0.672	&0.9	&0.929	&0.929 & & 0.734 & 0.956  & 0.601  & 0.635  & 0.658 & 0.659 \\
            & nnU-Net ($\gamma=0.9$)     &0.726	&0.945	&0.682	&0.898	&0.926	&0.926 & & 0.702 & 0.945  & 0.616  & 0.658  & 0.679 & 0.680 \\
            & nnU-Net ($\gamma=1.0$)     &0.686	&0.956	&0.673	&0.897	&0.925	&0.926 & & 0.675 & 0.956  & 0.605  & 0.651  & 0.675 & 0.675 \\
            & Swin UNETR                 &0.692	&0.956	&0.621	&0.892	&0.921	&0.921 & & 0.692 & 0.956  & 0.55  & 0.591 & 0.618 & 0.618 \\
            & Swin UNETR (FT)            &0.746	&0.956	&0.644	&0.881	&0.912	&0.913 & & 0.712 & 0.956  & 0.53  & 0.522 & 0.542 & 0.543 \\
            & HFF-Net                  &0.683	&0.934	&0.703	&0.9	&0.928	&0.928 & & 0.673 & 0.934  & 0.638  & 0.653  & 0.674 & 0.675 \\
            & Ensemble             &\textbf{0.723}	&\textbf{0.956}	&\textbf{0.689}	&\textbf{0.895}	&\textbf{0.923}	&\textbf{0.923} & & \textbf{0.71} & \textbf{0.956}  & \textbf{0.615}  & \textbf{0.629}  & \textbf{0.653} & \textbf{0.654} \\
            \bottomrule
        \end{tabular}%
    }
    \label{tab:ped-val}
\end{table}

\begin{table}[h!]
    \caption{\textbf{Quantitative results} on the test datasets of PED. 
    Our method ranks $1^{\text{st}}$ on the final test set leaderboard.
    Lesion-wise (LW) Dice coefficients and Normalized Surface Dice at 1.0 mm tolerance (NSD-1.0) were computed for enhancing tumor (ET), tumor core (TC), whole tumor (WT), non-enhancing tumor (NET), cystic components(CC), and edema (ED), respectively.}
    \renewcommand{\arraystretch}{1.15}  
    \centering
    \resizebox{1.0 \textwidth}{!}{%
        \begin{tabular}{@{}c c *{6}{c} c *{6}{c}@{}}
            \toprule
            \multirow{2}{*}{\textbf{Task}} & \multirow{2}{*}{~~~~\textbf{Statistic}} & \multicolumn{6}{c}{\textbf{Lesion-wise Dice $\uparrow$ }} & & \multicolumn{6}{c}{\textbf{Lesion-wise NSD-{1.0} $\uparrow$}} \\
            \cmidrule(lr){3-8} \cmidrule(lr){10-15}
            & & \textbf{CC} & \textbf{ED} & \textbf{ET} & \textbf{NET} & \textbf{TC} & \textbf{WT} & & \textbf{CC} & \textbf{ED} & \textbf{ET} & \textbf{NET} & \textbf{TC} & \textbf{WT} \\
            \midrule
            \multirow{2}{*}{\begin{tabular}[c]{@{}c@{}}PED\\(rank 1)\end{tabular}}
            & mean                  & \textbf{0.627} & \textbf{0.832} & \textbf{0.729} & \textbf{0.857} & \textbf{0.918} & \textbf{0.926} & & \textbf{0.637} & \textbf{0.831} & \textbf{0.776} & \textbf{0.823} & \textbf{0.833} & \textbf{0.844} \\
           & std  & 0.453 & 0.354 & 0.318 & 0.196 & 0.129 & 0.123 & & 0.446 & 0.354 & 0.322 & 0.203 & 0.217 & 0.203 \\
            \bottomrule
        \end{tabular}%
    }
    \label{tab:ped-test}
\end{table}

The results show that vanilla nnU-Net achieves an average Dice score of 83\% across six sub-regions. By tuning the initialization parameter $\gamma$, we obtain the highest average Dice score of 85.6\% at $\gamma=0.7$. The pre-trained Swin UNETR-FT outperforms its from-scratch counterpart by approximately 1\% in Dice score. Among all regions, ET remains the most challenging to segment. HFF-Net provides modest improvement for ET segmentation, reaching approximately 70\% Dice score. Our final ensemble combines nnU-Net ($\gamma=0.7$), Swin UNETR-FT, and HFF-Net, achieving Dice scores of 72.3\% (ET), 95.6\% (NET), 68.9\% (CC), 89.5\% (ED), 92.3\% (TC), and 92.3\% (WT).

Table \ref{tab:ped-test} presents the final results of our proposed ensemble method on the test datasets .

\subsection{Qualitative Results}
\label{sec:qualitative}

\begin{figure}[htb]
	\centering
	\includegraphics[width=\linewidth]{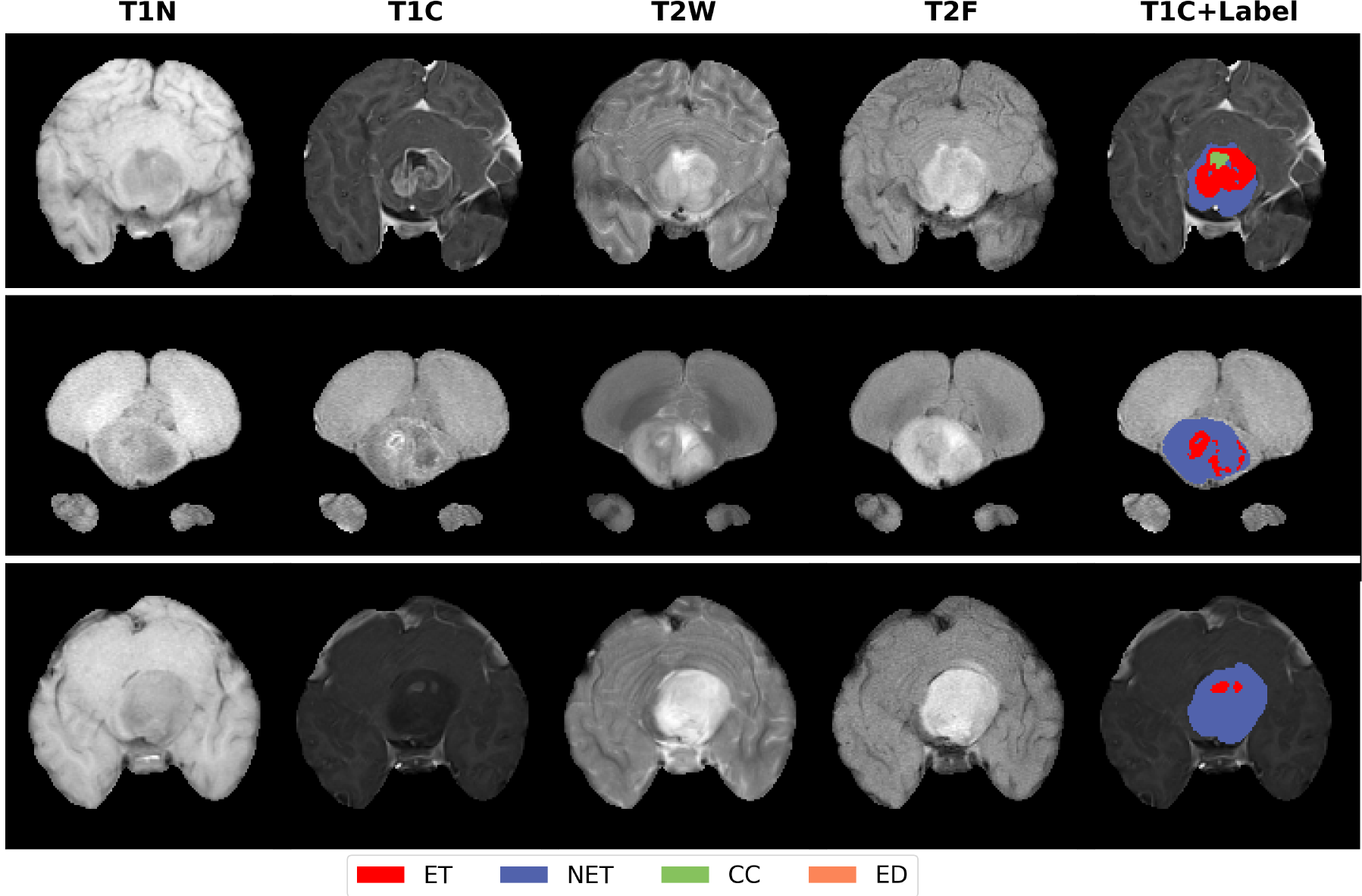}
	\caption{ \textbf{Quantitative results} of the final model on the validation set. The three selected examples correspond to BraTS-PED-00310-000, BraTS-PED-00315-000, and BraTS-PED-00318-000, respectively. }
	\label{fig:final_model}
\end{figure}

Fig.~\ref{fig:final_model} presents the qualitative segmentation results of the final ensemble model on validation data. The validation set underwent the same skull stripping and frequency domain decomposition pre-processing.

\section{Discussion}
\label{sec:discussion}

In this work, we propose a frequency-aware ensemble learning framework for pediatric brain tumor segmentation in the BraTS-PED 2025 challenge. By integrating nnU-Net, Swin UNETR, and HFF-Net, and leveraging techniques such as multi-scale parameter initialization, transfer learning, and frequency-domain decomposition, our approach achieves robust and accurate segmentation performance on challenging pediatric datasets. Experimental results demonstrate the effectiveness of the proposed ensemble strategy across multiple quantitative metrics.
Our final ensemble framework, comprising nnU-Net ($\gamma=0.7$), fine-tuned Swin UNETR, and HFF-Net, demonstrates strong generalization capability on unseen pediatric brain tumor data. 
The method achieves competitive Dice scores of 62.7\% (CC), 83.2\% (ED), 72.9\% (ET), 85.7\% (NET), 91.8\% (TC), and 92.6\% (WT) on the test datasets. 
More importantly, the proposed approach secures first place in the BraTS-PED 2025 Segmentation Challenge, underscoring its robustness and state-of-the-art performance in pediatric brain tumor segmentation.

In future work, we plan to further explore advanced model architectures and multi-modal data fusion techniques to enhance the clinical applicability of automated brain tumor segmentation.

\bibliographystyle{splncs04}
\bibliography{references} 

\end{document}